\documentclass[12pt]{iopart}
\usepackage{graphicx}
\begin{document}

\letter{Phase engineering of controlled entangled number states in a single component Bose-Einstein condensate in a double well}

\author{K W Mahmud\dag\footnote[3]{Author to whom correspondence should be addressed.}, H Perry\ddag and W P Reinhardt\dag\ddag}
\address{\dag\ Department of Physics, University of Washington, Seattle, WA 98195-1560, USA\\}
\address{\ddag\ Department of Chemistry, University of Washington, Seattle, WA 98195-1700, USA\\}

\date{\today}
\eads{kmahmud@u.washington.edu,perryh@u.washington.edu,rein@chem.washington.edu}

\maketitle

\begin{abstract}
We propose a model for the creation of entangled number states (Schr\"odinger cat states) of a Bose-Einstein condensate in a double well through simple phase engineering. We show that a $\pi$-phase imprinted condensate in a
double-well evolves, with a simultaneous change of barrier height, to number states with well defined and controlled entanglement. The cat state generation is understood in terms of the underlying classical phase space dynamics of a $\pi$-phase displaced coherent state put at the hyperbolic fixed point of the separatrix of a physical pendulum. The extremity and sharpness of the final cat state is determined by the initial barrier height and the rate at which it is ramped during the evolution. 
\end{abstract}

The superposition principle of quantum mechanics, when applied to
macroscopic objects, gives rise to seeming paradoxes as  highlighted by
Schr\"odinger in his famous cat discussion~\cite{schrodinger}. Unlike
superpositions of different spin states of an electron, a classical object
such as a room temperature cat cannot be found in a superposition of two
macroscopically distinct states. However, carefully controlled experiments have been performed to
create such superposition states of mesoscopic atomic and
condensed matter systems~\cite{friedman1,cat2}. Due to the macroscopic
nature of its wavefunction, the Bose-Einstein condensate (BEC) should be an
ideal system in which to create such macroscopic superposition states. There have
been several proposals to create cat states with BECs, although none have
been demonstrated experimentally~\cite{cirac1,savage1,burnett1,ruostekoski1,ruostekoski2}. In this Letter we propose a novel model for the creation of controlled entangled number states of a BEC in a double well via phase imprinting on the part of the condensate in one of the wells followed by a change of barrier height. When properly implemented this results in a state of the form

\begin{equation}
|\Psi\rangle=\frac{1}{\sqrt{2}}\left(|n_L,N-n_L\rangle+|N-n_L,n_L\rangle\right) 
\end{equation}
where $|n_L,n_R\rangle$  denotes a state with $n_{L}$ particles in the left well, $n_{R}$ in the right well, and the total number of particles is $N=n_{L}+n_{R}$. Unlike in related proposals for spinor condensates~\cite{savage1,burnett1}, we can use the barrier height to control the squeezing of the initial BEC ground state followed by a continuous change of barrier height to control the value of $n_{L}$ ($n_{L}=0,1,2...N$) of the final cat state and thereby control both the extremity (the value of $n_{L}$) and the sharpness (the spread around $n_L$) of the cat state. An extreme cat state would correspond to $n_{L}=0$ or $N$. Ground state number squeezing with variable barrier height in double and multi-well systems has been discussed and observed by many authors~\cite{spekkens1,kasevich2,mott1}.

\begin{figure}
\begin{center}
\includegraphics[width=8.2cm]{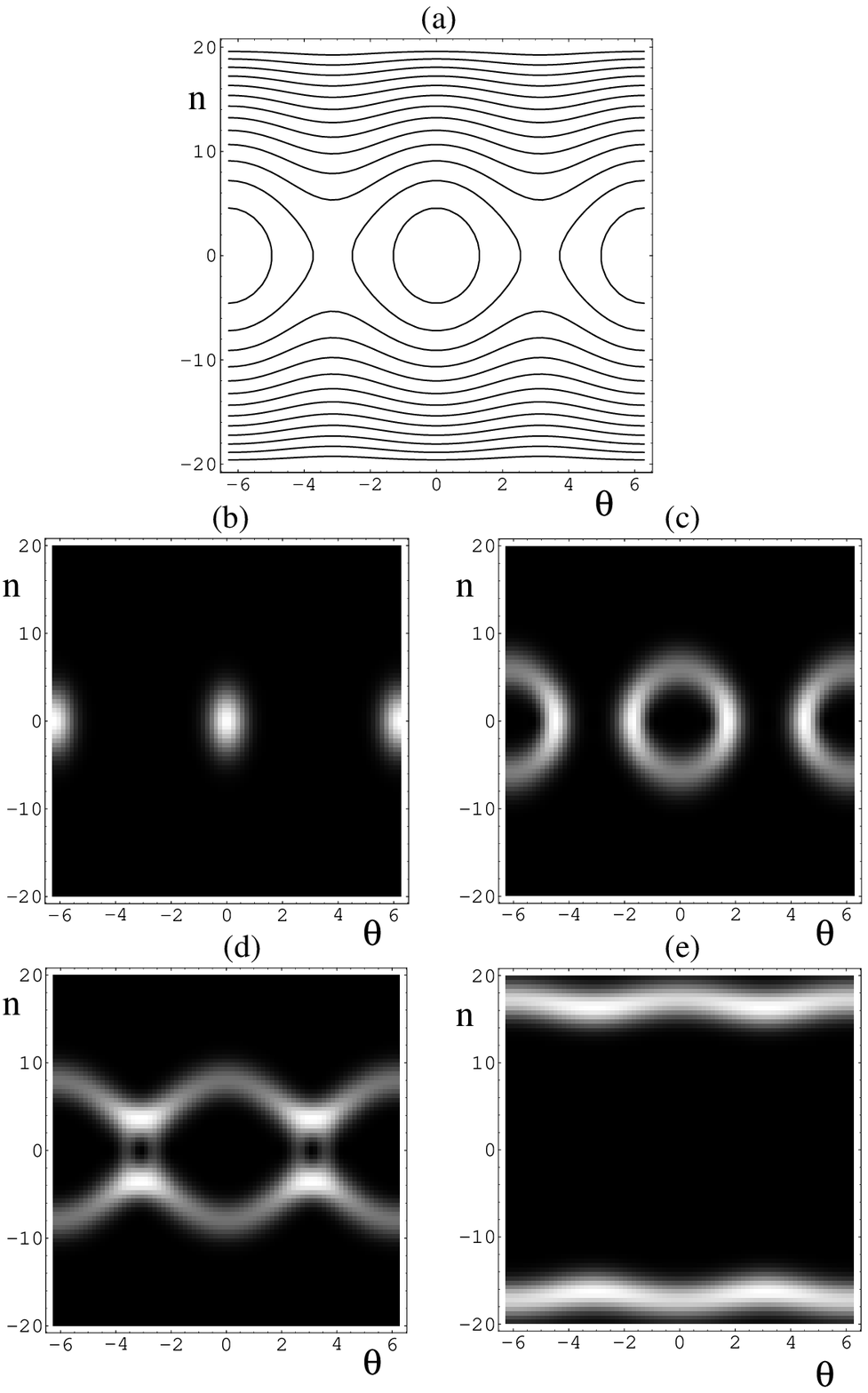}
\end{center}
\caption{
Comparison of the classical nonrigid pendulum phase space and Husimi probability distributions for different energy eigenstates for 40 particles. Shown are (a) classical energy contours. Husimi projections for (b) the ground state which is a minimum uncertainty wave-packet centered at the origin, (c) the 6th state which is harmonic-oscillator like and the analog of pendulum librations, (d) the 12th state which is a quantum separatrix state separating the libration and rotor states, (e) the 35th state which is analogous to a superposition of classical pendulum rotor motions in two opposite directions. The creation of such macroscopic superposition states with a BEC is possible with phase engineering.}
\label{fig:regularcompL}
\end{figure}

The physical motivation for what is demonstrated here is classical:
a wave-packet placed at the unstable fixed point of a classical phase space separatrix will bifurcate
if allowed to time evolve. The appropriate separatrix in this case is that
dividing bounded from rotor motion of a simple physical pendulum. It has
long~\cite{anderson2} been noted that a double well quantum fluid may be
modeled as a physical pendulum, an analysis more recently extended in Ref.~\cite {smerzi1,smerzi2}. Modeling a repulsive BEC in a double well in the two
mode approximation~\cite{spekkens1,milburn1}, the coefficients of the exact
Fock space eigenstates may be projected into a semi-classical ($n,\theta $)
phase space via the Husimi probability distribution~\cite{husimi,husimiReview1}, whereby
the pendulum analogy is immediately seen to be a property of the exact
quantum eigenfunctions, as illustrated in Fig.~\ref{fig:regularcompL}. The low energy states correspond to  oscillatory motion of the pendulum, and
the higher-lying states to superpositions of pendulum rotor motions in two
opposite directions, these being macroscopic superposition states. The
ground state in this model is a Gaussian centered at the origin  in ($n,\theta$) space with number squeezing controlled by the barrier height~\cite{spekkens1}. A displacement in phase by $\pi $, as may be created by phase
shifting the single particle wavefunction in one of the wells, shifts the
ground state wave-packet to the unstable fixed point. As expected, the
subsequent autonomous dynamics of the shifted packet lead to bifurcation
and production of macroscopic superposition states. These may be further
controlled, leading to sharp number entangled states with a chosen value of 
$n_{L}$, as in Eqn. (1), by choice of the initial state and barrier ramping, as discussed below.  Further details of the quantum
phase space picture and its applications are presented elsewhere~\cite{kmahmud4}.
      
The many-body Hamiltonian for a system of $N$ weakly interacting bosons in an external potential $V({\bf r})$ in second quantization, is given by      
\begin{eqnarray}
\hat{H}=\int d{\bf r}\hat{\Psi}^{\dagger}({\bf r})\left[-\frac{\hbar^2}{2m}\nabla^2+V({\bf r})\right]\hat{\Psi}({\bf r})\nonumber\\
+\frac{g}{2}\int d{\bf r}\hat{\Psi}^{\dagger}({\bf r})\hat{\Psi}^{\dagger}({\bf r})\hat{\Psi}({\bf r})\hat{\Psi}({\bf r})
\label{eqn:Nhamil}
\end{eqnarray}
where $\hat{\Psi}({\bf r})$ and $\hat{\Psi}^{\dagger}({\bf r})$ are the bosonic annihilation and creation field operators, $m$ is the particle mass, and  $g=\frac{4\pi a_s \hbar^2}{m}$ where $a_s$ is the s-wave scattering length.

Many authors have studied the double-well condensate using the two-mode approximation~\cite{spekkens1,milburn1}. We use the model introduced by Spekkens and Sipe~\cite{spekkens1}. The exclusion of the nonlinear tunneling terms in this model gives rise to the Bose-Hubbard model~\cite{bosehubbard1}. The full two-mode Hamiltonian is    
\begin{eqnarray}
\hat{H}=\epsilon_{LL}\hat{N_{L}}+\epsilon_{RR}\hat{N_{R}}+(\epsilon_{LR}+gT_{1}(\hat{N}-1))(a^{\dagger}_{L}a_{R}+a^{\dagger}_{R}a_{L})\nonumber\\
+\frac{gT_0}{2}(\hat{N}^2_{L}+\hat{N}^2_{R}-\hat{N})\nonumber\\
+\frac{gT_2}{2}(a^{\dagger}_{L}a^{\dagger}_{L}a_{R}a_{R}+a^{\dagger}_{R}a^{\dagger}_{R}a_{L}a_{L}+4\hat{N_{L}}\hat{N_{R}})
\label{eqn:sipeHamil}  
\end{eqnarray}           
where $\hat{N_{L}}=a^{\dagger}_{L}a_{L}$, $\hat{N_{R}}=a^{\dagger}_{R}a_{R}$, $\hat{N}=\hat{N_{L}}+\hat{N_{R}}$ and
\begin{equation}
\epsilon_{ij}=\int d{\bf r}\phi_{i}({\bf r})\left(-\frac{\hbar^2}{2m}\nabla^2+V({\bf r})\right)\phi_{j}({\bf r})\nonumber\\
\end{equation}
where i,j=L,R.
\begin{eqnarray}
T_0=\int d{\bf r}\phi^4_{L}({\bf r});T_1=\int d{\bf r}\phi^3_{L}({\bf r})\phi_{R}({\bf r});
T_2=\int d{\bf r}\phi^2_{L}({\bf r})\phi^2_{R}({\bf r}).
\end{eqnarray}

$\epsilon _{LL}$ and $\epsilon _{RR}$ are the energies of a single particle
in the left and right wells, each is multiplied by the number operators $N_{L,R}$ counting the number of particles in each well; $\epsilon _{LR}$ is
the single particle tunneling amplitude which appears as a coefficient of
operators that allow a single particle to hop from one well to the other; $T_{0}$ is the nonlinear mean-field energy in each well and $T_{1,2}$ are
nonlinear tunneling matrix elements.

To simplify a theoretical study of this two-mode Hamiltonian, we make a one parameter approximation of the single particle energies and the tunneling matrix elements:
\begin{eqnarray}
\epsilon_{LL}=\epsilon_{RR}=T_0=1;\epsilon_{LR}=T_1=-e^{-\alpha};T_2=-e^{-2\alpha}.
\end{eqnarray}
This parametrization allows a simple study of continuous change in the linear and non-linear tunneling through variation of a single parameter $\alpha$, which is associated with the barrier height and width. In our computations with this model we ignore the $T_2$ term which scales as $e^{-2\alpha}$. 

\begin{figure}
\begin{center}
\includegraphics[width=8.2cm]{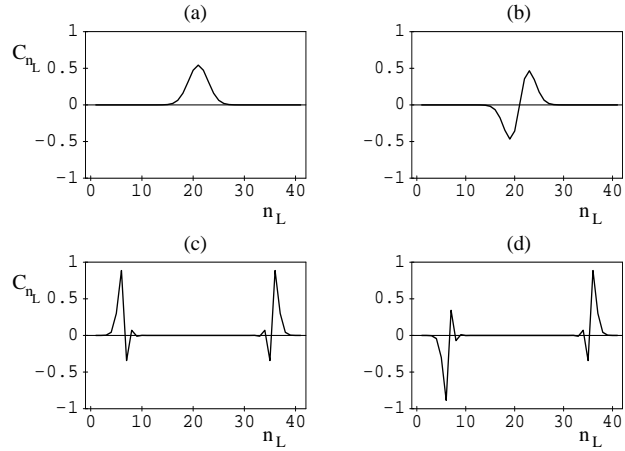}
\end{center}
\caption{Fock state coefficients for 40 particles for (a) the ground state, (b) the first excited state, (c) the 30th state and (d) the 31st state. Low lying states are similar to harmonic oscillator wavefunctions, whereas the higher lying states are cat-like.}
\label{fig:coeffsL}
\end{figure}

Taking into account the conservation of the total number of particles, the state space for the system can be spanned in $N+1$ Fock state basis $|n_L,N-n_L\rangle$, $n_L=0,1,...N$, where $n_L$ particles occupy the single particle state in the left well and $N-n_L$ particles occupy the right well. The state vector is a superposition of all the number states
\begin{equation}
|\Psi\rangle=\sum_{n_L=0}^{N}c^{(i)}_{n_L}|n_L,N-n_L\rangle
\label{eqn:basis}
\end{equation} 
where 
\begin{equation}
|n_L,N-n_L\rangle=\frac{(a^{\dagger}_{L})^{n_L}}{\sqrt{n_L}}\frac{(a^{\dagger}_{R})^{N-n_L}}{\sqrt{N-n_L}}|vac\rangle.
\end{equation}
Finding the eigenvalues and eigenvectors of the model Hamiltonian in the
Fock basis can be easily accomplished by diagonalizing a $(N+1)\times(N+1)$
tridiagonal matrix. Fig.~\ref{fig:coeffsL} panels (a) and (b) show the
coefficients of the eigenvectors for the two lowest lying states for 40
particles for $\alpha=4$. Panels (c) and (d) show the 30th and the 31st eigenstates. The
lowest lying states clearly have harmonic oscillator like wavefunctions in
Fock space. On the other hand the higher lying states are cat-like, i.e. superpositions of states with most particles in the left and right wells.

The characteristics of these states can be understood in terms of the motion
of a momentum-shortened physical pendulum. Anderson suggested that the dynamics of the Josephson effect can be understood in terms of a physical pendulum~\cite{anderson2}; similarly the mean-field dynamics of a Bose-Einstein condensate in a double well is well
described by a momentum-shortened physical pendulum, which has novel phase
space characteristics~\cite{smerzi1,smerzi2,kmahmud4,penna2}. It is natural
to ask what aspects of the mean-field phase space properties are contained
in the exact quantum treatment. Phase space formulation of quantum mechanics
is extremely useful in studying many aspects of quantum optics, collision
theory and quantum chaos, and this is the approach we take to study the
double well. The Husimi probability distribution~\cite{husimi,husimiReview1} can be used to project, in a squeezed coherent state representation, the classical phase space properties from stationary quantum wavefunctions. For a BEC in a double well, the phase difference $\theta $=$\theta _{L}$-$\theta_{R}$ and the number difference $n=\frac{n_{L}-n_{R}}{2}$ are the conjugate variables analogous to q and p. In this (n,$\theta$) representation a probability distribution function can be defined as 
\begin{equation}
P_{j}(n,\theta)=\vert\langle\theta+in|\Psi_{j}\rangle\vert^2
\end{equation}
where
\begin{equation}
\langle\theta+in|\Psi_{j}\rangle=\frac{1}{(\kappa \pi)^4} \sum_{n'=-N/2}^{N/2}c^{j}_{n'}\exp\lbrack i\theta n'-\frac{(n'-n)^2}{2 \kappa}\rbrack.
\label{eqn:husimi}
\end{equation}
Here $n'=\frac{n_{L}-n_{R}}{2}$, rather than being the simpler left particle counter, and $c_{n'}$ are the corresponding Fock-state coefficients. The Q-function in quantum optics is a special case of the Husimi distribution
function where $\kappa $=$\omega ,$  $\omega $ being the frequency of a
coherent state Gaussian wave-packet~\cite{husimiReview1}. The
`coarse-graining' parameter $\kappa $ determines the relative resolution
in the conjugate number-phase (n,$\theta$) space.

Fig.~\ref{fig:regularcompL} panels (b), (c), (d) and (e) show the density
plots for Husimi probability distributions for 40 particles for the ground
state, 6th, 12th and 35th states respectively. Panel (a) shows the classical
energy contours that are also the classical trajectories. As is evident from
the panels, the ground state is a minimum uncertainty wave-packet in both
number and phase that is centered at the origin, the
harmonic-oscillator-like low lying excited states are the analog of pendulum
librations, and the higher lying cat-like states are the analog of rotor
motions, with a clear signature of the quantum separatrix state where the
libration and rotation states separate. Note that the Husimi distribution of
the 35th state shown in panel (e) is a superposition of most particles in
the left and right wells. In the classical phase space this corresponds to
two trajectories above and below the separatrix that denote rotor motions of
a physical pendulum in two opposite directions. The creation of such
macroscopic superposition states with BEC is possible with simple phase
engineering. It is clear that near classical behavior is observed for quite small values of $N$. Larger $N$ further sharpens the classical correspondence.

Writing phases on part of a condensate is experimentally feasible via
interaction with a far off-resonance laser. This method has been used to
generate dark solitons and measure their velocities due to a phase differential~%
\cite{billsoliton,darksoliton2}. Mathematically, such a method corresponds to
multiplying the coefficient of each of the Fock state basis of an eigenstate with $e^{-in_{L}\theta}$, where $|n_{L}\rangle $ is the corresponding Fock state, and $\theta $ is the phase offset for particles in the left well. By $\pi $ phase imprinting the
condensate in one well, the ground state centered at the origin (0,0) in
phase space is displaced to the unstable equilibrium point (0,$\pi $) on
the separatrix. Using exact quantum time evolution within the framework of
the two mode model, the resulting quantum wave-packet bifurcates as
expected.  If the barrier is raised as discussed below, the wave-packet is permanantly split, resulting in a superposition of two classical rotor states. With the
appropriate ramping of the potential barrier and proper tuning of the values
of the parameters in the model we can vary the sharpness and extremity of
the cat states.

\begin{figure}
\begin{center}
\includegraphics[width=8.2cm]{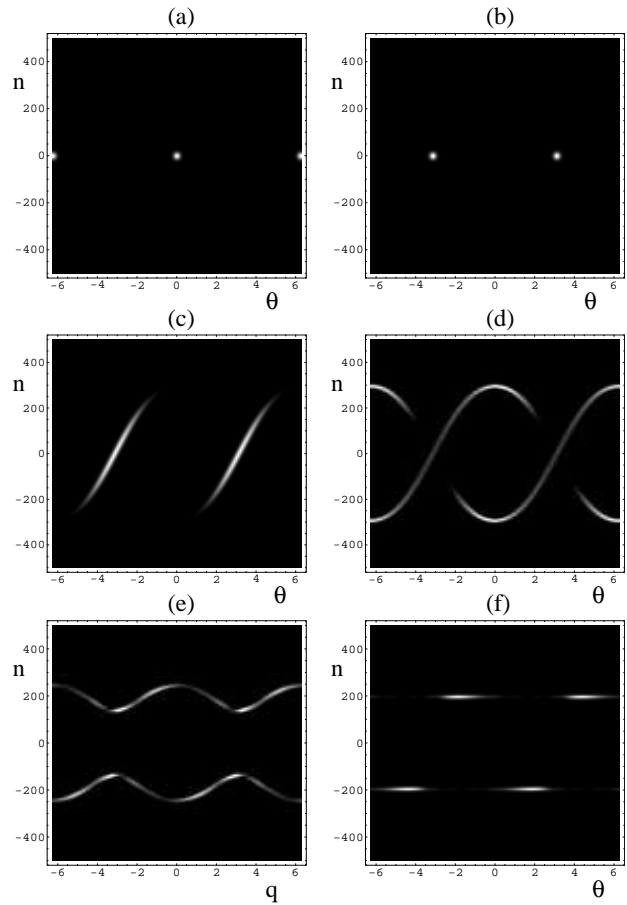}
\end{center}
\caption{
Shown is the evolution to a cat state in Husimi projection space. (a) The ground state at t=0, (b) the $\pi$-phase imprinted ground state at the hyperbolic fixed point, (c) at t=0.01 the wave-packet is bifurcating along the separatrix, (d) at t=0.016 it continues to move along the separatrix, (e) at t=0.4 the states become trapped as we increase the barrier, and (f) at t=2.3 a sharply peaked cat state is obtained.
}
\label{fig:cathusimiL}
\end{figure}

\begin{figure}
\begin{center}
\includegraphics[width=8.2cm]{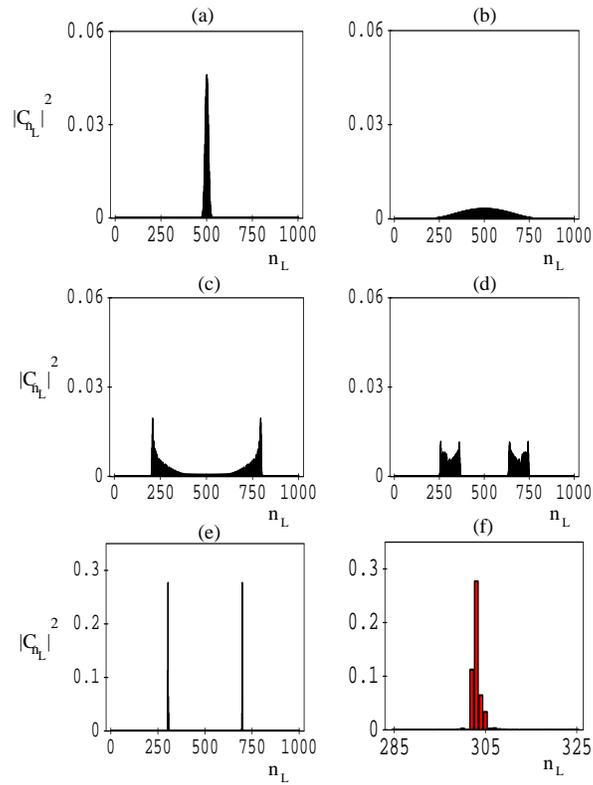}
\end{center}
\caption{
Shown is the evolution to a cat state in Fock space at the same time instants as the previous figure: (a) t=0, (b) t=0.01, (c) t=0.016, (d) t=0.4, (e) t=2.3, (f) is a magnified version of (e), showing the nonvanishing Fock state coefficients. Note that for clarity the probability amplitudes in the vertical axes have different scalings for the panels.
}
\label{fig:catEvolL}
\end{figure}

\begin{figure}
\begin{center}
\includegraphics[width=10.5cm]{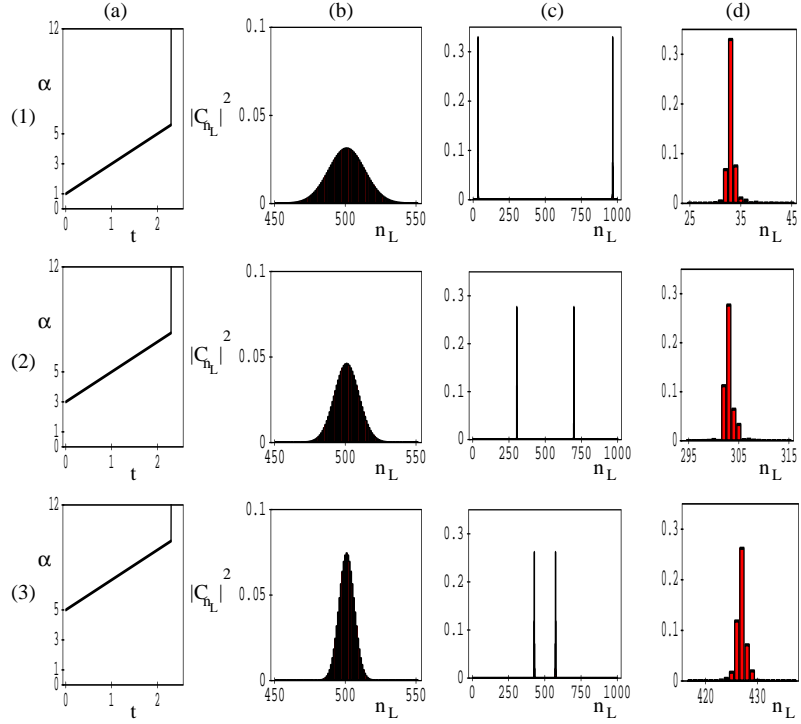}
\end{center}
\caption{
Shown are the cat states with different initial heights of the barrier and therefore different initial squeezings of the BEC ground state, but the same ramping of the potential. Row (1) shows the states where $\alpha=1+2t$: (a) the parameter $\alpha$ as a function of time, (b) the ground state, (c) the final cat state, and (d) a magnified view of the Fock-state coefficients. Rows (2) and (3) show the results for $\alpha=3+2t$ and $\alpha=5+2t$ respectively. The initial barrier height controls the extremity of the cat states. Note that for clarity the axes in the panels have different scalings. 
}
\label{fig:catcompareL}
\end{figure}

As an example, in
Figs.~\ref{fig:cathusimiL} and~\ref{fig:catEvolL} we show how a cat state with 1000 particles is generated. Fig.~\ref{fig:cathusimiL} shows the evolution in phase space using Husimi projections
- (a) the ground state, (b) a $\pi$-phase imprinted state, (c) and (d) show subsequent evolution in the process of bifurcating the state; further evolution along with a change of barrier totally splits and traps the state above and below the separatrix, as shown in (e), finally giving rise to cat state in (f). Shown in Fig.~\ref{fig:catEvolL} is the evolution in the Fock state basis. A close view of the coefficients, shown in (f), shows that
it is a rather sharply peaked entangled number state. The parameters used
for this example are $g=T_0=1$, $\epsilon_{LR}=T_1=-e^{-\alpha}$ where
the barrier height is ramped up in time as $\alpha=3+2t$. When a cat state
is reached the barrier is suddenly raised to essentially halt the evolution.
With different initial barrier heights and the same ramping of
the potential, the extremity of the cat states can be tuned. Examples are shown
in Fig.~\ref{fig:catcompareL} where the different values of the barrier
heights are $\alpha=1+2t$, $\alpha=3+2t$ and $\alpha=5+2t$ for rows (1), (2) and (3) respectively. The columns show: (a) the barrier height and the ramping,
(b) the respective ground state, (c) the final cat state at the end of the ramping, and (d) a close view of the coefficients for the final state. As is evident from the pictures, the initial squeezing of the ground
state determines the extremity of the final cat state. The rate at which the barrier is ramped determines the sharpness. It is important to be
able to tune to less extreme cat states as such states are more robust to
loss and decoherence.

In conclusion, we have shown that a $\pi$-phase imprinted BEC in a double
well evolves to an entangled number state (Schr\"odinger cat state) which is a
superposition of most particles being in the left and right wells simultaneously. Initial squeezing of the ground state and a ramping of the barrier height help tune the extremity and sharpness of the final cat state. A semi-classical phase space analysis of the full quantum problem reveals the
similarity with a classical pendulum phase space, thus providing a
visual explanation of cat state generation through motions along the
separatrix. Making condensates in a double-well, phase imprinting and 
tuning the barrier are viable current experimental
techniques. However, decoherence mechanisms will put severe constraints on
the size or extremity of cat states that can be produced in a
laboratory. Such challenges will make the attainment of cat states with a BEC
interesting and important.

%%%%%%%%%%%%%%%%%%%%%%%%%%%%%%%%%%%%%%%%%%%%%%%%%%%%%%%%%%%%%%%%%%%%%%%%%%%%%%%%%%%

We would like to thank Sarah B. McKinney for discussions and computational support. This work was supported by NSF grant PHY-0140091.

%%%%%%%%%%%%%%%%%%%%%%%%%%%%%%%%%%%%%%%%%%%%%%%%%%%%%%%%%%%%%%%%%%%%%%%%%%%%%%%%%%

%

\end{document}